\documentclass{amsart}
\usepackage[T1]{fontenc}
\usepackage{amsfonts}
\usepackage{amssymb}
\usepackage{multirow} 
\usepackage{longtable} 
\usepackage{array} 
\usepackage{makecell} 
\usepackage{threeparttable}
\usepackage{tablefootnote}
\usepackage[colorlinks=true,linkcolor=blue,allcolors=blue
]{hyperref}
\usepackage{amsmath,blkarray,booktabs,bigstrut}
\usepackage{tikz}
\usepackage{tikz-cd}
\usepackage{graphicx}
\usepackage{dutchcal}
\usepackage{algorithm}
\usepackage{algpseudocode}
\usepackage{soul}
\usepackage{cleveref}
\usepackage{float}
\usepackage[left=3.3cm,right=3.3cm, top=3.2cm,bottom=3.2cm]{geometry}
\numberwithin{equation}{section}

\theoremstyle{plain}
  \newtheorem{theorem}[subsection]{Theorem}

\theoremstyle{definition}

\newif\ifTYshow
\newif\ifTYRshow

\ifTYshow
  
\else
  
\fi

\usepackage[normalem]{ulem} 
  
\begin{document}
\title[WASSERSTEIN TARGET DIAGRAM]{Wasserstein Target Diagram for Model Evaluation}

\author{Tiantian Yang}
 \address{Department of Mathematics and Statistical Science, University of Idaho, Moscow, ID, USA, 83844}
 \email{tyang@uidaho.edu}

\author{Dongwei Chen}
 \address{Department of Mathematics and Statistical Science, University of Idaho, Moscow, ID, USA, 83844}
 \email{dongweic@uidaho.edu}

\begin{abstract}
The target diagram is often used in geosciences, which provides a visualization tool for model evaluation and shows the temporal or spatial matching of data. However, its construction requires paired data of equal sample sizes, limiting its application to model comparisons involving unpaired samples or probability distributions. In this study, we propose the Wasserstein target diagram based on the 2-Wasserstein distance from optimal transport. Unlike the original target diagram, the Wasserstein target diagram does not require paired data of equal sizes and evaluates models from the probability distribution perspective. We further establish that the two target diagrams coincide when the given paired data are comonotone, which is common in climate science, such as cumulative precipitation and solar radiation. Through two simulation studies and an application to wind speed distribution fitting, we demonstrate that the proposed diagram can detect the probability distribution discrepancies that the original target diagram may not distinguish. The Wasserstein target diagram therefore lifts the model evaluation to the probability level and provides a complementary approach to estimating the agreement in probability distributions.
\end{abstract}


\maketitle

\section{Introduction}
Evaluating the agreement between observations and model predictions is an important task in statistical modeling and scientific applications. Common evaluation measures often use statistical metrics to quantify the model prediction performance. For example, the bias of mean values measures systematic overestimation or underestimation, while the root mean squared error (RMSE) summarizes the overall magnitude of prediction errors. In the last two decades, the Taylor diagram \cite{taylor2001summarizing} and the target diagram \cite{jolliff2009summary}, which both visualize the evaluation metrics in a single diagram, have been widely used in model evaluation. The Taylor diagram uses the law of cosines to visualize the evaluation metrics, including correlation coefficient, centered root mean squared error (CRMSE), and variances of observed and predicted data. However, the Taylor diagram ignores the mean values. To solve this problem, the authors in \cite{jolliff2009summary} introduced the target diagram by decomposing RMSE into bias of mean values and CRMSE, which satisfy the Pythagorean theorem.  
The mathematical construction of the target diagram is as follows.  

\begin{figure*}[ht]
\centering
\includegraphics[scale=0.6]{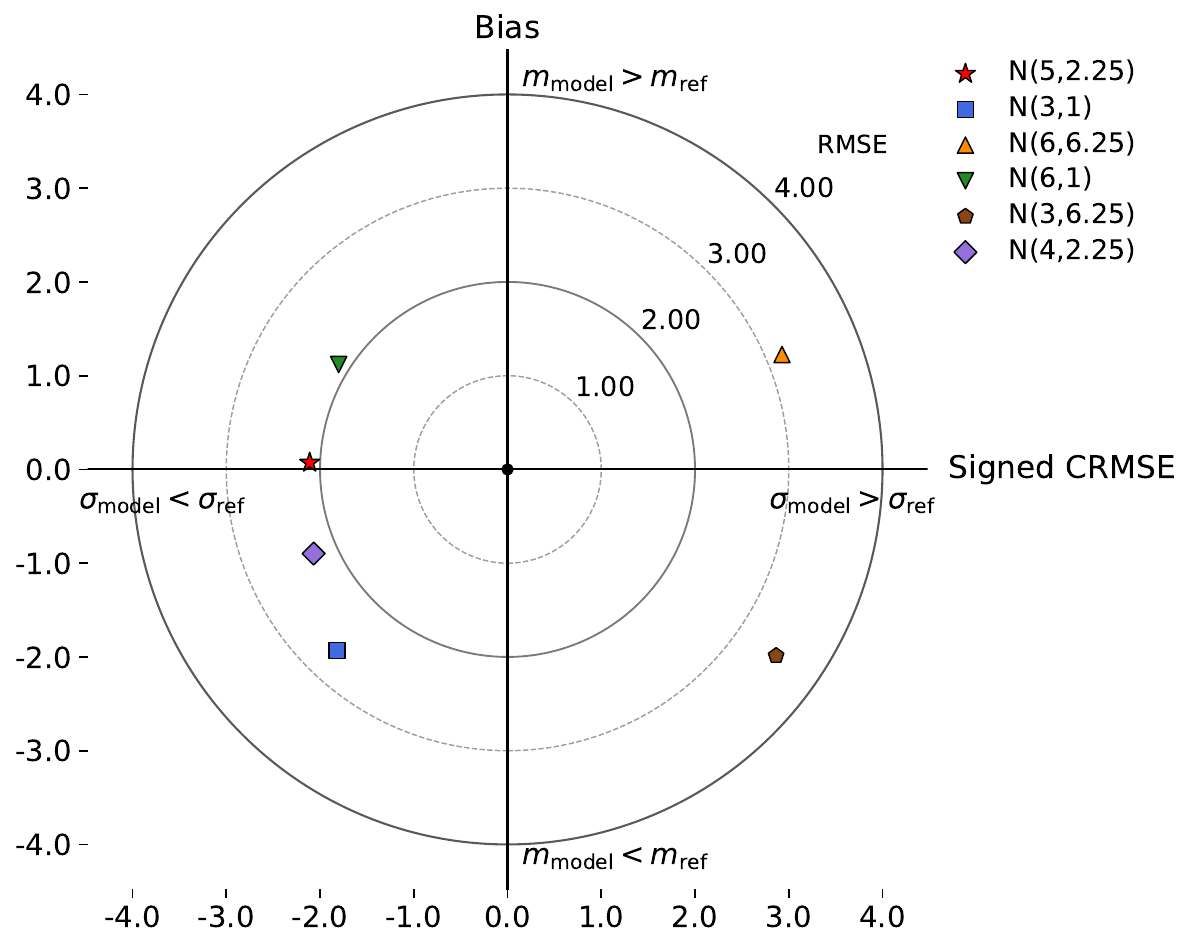}
\caption{Target Diagram Example. The reference sample consists of $n=1{,}000$ observations generated from a normal distribution $N(5,2.25)$, with mean $4.940654$ and standard deviation $1.501181$. Predicted data are generated from six different distributions. All samples have equal size ($n=1{,}000$) and are paired with the reference. The horizontal coordinate represents the signed CRMSE, the vertical coordinate represents bias, and radial distance represents RMSE. Positive (negative) horizontal coordinates indicate that predicted data have larger (smaller) standard deviations than the reference. Positive (negative) vertical coordinates indicate that predicted data have larger (smaller)  means than the reference.}
\label{fig:EqualLength_TargetSimulation}
\end{figure*}

Given predicted data $\{x_i\}_{i=1}^N$ and observed data $\{y_i\}_{i=1}^N$, the RMSE is given as
\begin{equation*}
\operatorname{RMSE}^2
=\frac{1}{N}\sum_{i=1}^N(x_i-y_i)^2 =\frac{1}{N}\sum_{i=1}^N\left(\bar{x}-\bar{y}+(x_i-\bar{x})-(y_i-\bar{y})\right)^2,
\end{equation*}
where $\bar{x}$ and $\bar{y}$ are the means of $\{x_i\}_{i=1}^N$ and $\{y_i\}_{i=1}^N$, respectively. 
Let $x_i'=x_i-\bar{x}$ and $y_i'=y_i-\bar{y}$ be the centered data of $\{x_i\}_{i=1}^N$ and  $\{y_i\}_{i=1}^N$. Then
\begin{equation*}
\begin{aligned}
\operatorname{RMSE}^2
&=\frac{1}{N}\sum_{i=1}^N(\bar{x}-\bar{y})^2
+\frac{2}{N}\sum_{i=1}^N(\bar{x}-\bar{y})(x_i'-y_i') 
+\frac{1}{N}\sum_{i=1}^N(x_i'-y_i')^2\\
&=\operatorname{Bias}^2+\operatorname{CRMSE}^2,
\end{aligned}
\end{equation*}
where $\operatorname{Bias}=\bar{x}-\bar{y}$ and $\text{CRMSE}$ is the RMSE of centered data  given by
\begin{equation*}
\operatorname{CRMSE}^2
=\frac{1}{N}\sum_{i=1}^N(x_i'-y_i')^2,
\end{equation*}
and the cross term vanishes since
\begin{equation*}
\begin{aligned}
\sum_{i=1}^N(x_i'-y_i')
&=\sum_{i=1}^N(x_i-\bar{x})
-\sum_{i=1}^N(y_i-\bar{y})=0.
\end{aligned}
\end{equation*}
Given $a, b \in \mathbb{R}$, define
\begin{equation*}
\text{sign}(a-b)=
\begin{cases}
1, & a> b,\\ 
0, & a=b,  \\
-1, & a<b.
\end{cases}
\end{equation*}
Let $\sigma_x$ and $\sigma_y$ be the sample standard deviations of $\{x_i\}_{i=1}^N$ and  $\{y_i\}_{i=1}^N$, respectively. Then a point in the target diagram with respect to $\{x_i\}_{i=1}^N$ and  $\{y_i\}_{i=1}^N$ has coordinates $(s, t)$ where
\begin{equation*}
s=\text{sign}(\sigma_x-\sigma_y)\cdot\operatorname{CRMSE} \ \text{and} \ 
t=\operatorname{Bias}=\bar{x}-\bar{y}.
\end{equation*}
When $\sigma_x\neq\sigma_y$, $\text{sign}(\sigma_x-\sigma_y) = \pm 1$, and we have the following Pythagorean theorem
\begin{equation}\label{eqn:target_decomposition}
\operatorname{RMSE}^2 =\operatorname{CRMSE}^2+\operatorname{Bias}^2
=s^2+t^2.
\end{equation}
Therefore,  the radial distance from the point $(s,t)$ to the origin in the target diagram represents RMSE. 
The sign of the horizontal coordinate indicates whether the predictions have a larger or smaller standard deviation than the observations, while the sign of the vertical coordinate shows whether the predictions have a larger or smaller mean value than the observations. Therefore, the target diagram represents the spatial or temporal matching of predicted data to observations, and allows multiple models to be compared within a common coordinate system. The closer a model is to the origin in the target diagram, the smaller its RMSE and the better its performance. An illustration of the target diagram is presented in Figure~\ref{fig:EqualLength_TargetSimulation} using six prediction samples of equal size generated from different distributions.

The target diagram requires paired observed and predicted data of equal length, since RMSE and CRMSE are based on pointwise differences. Consequently, it cannot be directly applied when the two samples have different sizes or only observed data exist. Furthermore, in many applications, the objective is to compare probability distributions rather than paired observations, for example, how to choose the best model when observed data are fitted to different probability distributions. In such settings, RMSE is not directly applicable without introducing an artificial pairing.  
Motivated by this insight, we lift the target diagram into the probability distribution level and
propose the Wasserstein target diagram to evaluate agreement in probability distributions.  Relevant studies on the Taylor and Wasserstein Taylor diagrams can be found in \cite{chen2026taylor}.
The Wasserstein distance, originating from the optimal transport theory, provides a natural approach to comparing probability distributions ~\cite{figalli2021invitation}. In particular, for two probability measures $\mu$ and $\nu$ on $\mathbb{R}$ with finite second moments,  the 2-Wasserstein distance ($W_2$) satisfies the decomposition
\begin{equation}\label{eqn:wtarget_decomposition}
W_2^2(\mu,\nu)=(m_\mu-m_\nu)^2+W_2^2(\widetilde{\mu},\widetilde{\nu}),
\end{equation}
where $m_\mu$ and $m_\nu$ are the distribution means, and $\widetilde{\mu}$ and $\widetilde{\nu}$ denote the corresponding centered distributions. Similar to the role of Equation \ref{eqn:target_decomposition} for the target diagram, Equation \ref{eqn:wtarget_decomposition} supports the construction of the Wasserstein target diagram.

The main contributions of this work are as follows. First, we develop the Wasserstein target diagram as a visual representation of the decomposition in Equation \ref{eqn:wtarget_decomposition} and evaluate models from the probability distribution perspective. It can compare unpaired samples and samples of different sizes, and detect distributional discrepancies beyond the bias.  Second, in Theorem \ref{thm:coincidence}, we show that the target diagram and the Wasserstein target diagram coincide if and only if the given data are comonotone, where the data are taken as equally weighted empirical distributions in the Wasserstein target diagram. Finally, we demonstrate the proposed approach through two simulation
studies with equal and unequal sample sizes and an application to the distribution fitting of wind speed. 

The rest of the paper is organized as follows. Section~\ref{sec:prelim} introduces the preliminaries of optimal transport and the Wasserstein distance. Section~\ref{sec:wtarget} presents the construction and theoretical properties of the Wasserstein target diagram. Section~\ref{sec:simulation} illustrates the proposed diagram through simulation studies and a real-world data application in wind speed. Finally, Section~\ref{sec:discussion} provides the discussion and conclusion.

\section{Preliminaries}\label{sec:prelim}

Let $\mathcal{P}(\mathbb{R})$ denote the set of Borel probability measures on the real line $\mathbb{R}$, and let $\mathcal{P}_2(\mathbb{R})\subset\mathcal{P}(\mathbb{R})$ denote the set of Borel probability measures with finite second moments, 
i.e.,
\begin{equation*}
\mathcal{P}_2(\mathbb{R})=\left\{\mu\in\mathcal{P}(\mathbb{R}):\int_{\mathbb{R}}|x|^2\,d\mu(x)<\infty\right\}.
\end{equation*}
Given $\mu \in  \mathcal{P}(\mathbb{R})$ and a Borel measurable map $T: \mathbb{R} \rightarrow \mathbb{R}$, the \emph{pushforward} of $\mu$ by $T$, denoted by $T_\#\mu$, is defined by
$$T_{\#} \mu (B) : = \mu \big (T^{-1}(B) \big ) \ $$
for every Borel set $B\subseteq\mathbb{R}$.
For $\mu\in\mathcal{P}_2(\mathbb{R})$, its mean $m_\mu$ and standard deviation $\sigma_\mu$ are given as
\begin{equation}
m_\mu:=\int_{\mathbb{R}}x d\mu(x) \ \text{and} \ \sigma_\mu:=\left[\int_{\mathbb{R}}(x-m_\mu)^2d\mu(x)\right]^{1/2}.
\label{mean_mu}
\end{equation}
The \emph{centered measure} of $\mu$ is defined as $\widetilde{\mu}:=(\tau_\mu)_\#\mu$ where $\tau_\mu(x)=x-m_\mu$.

It is known that the 2-Wasserstein distance ($W_2$) from optimal transport is a metric on $\mathcal{P}_2(\mathbb{R})$  and is often used to quantify the similarities and discrepancies between probability distributions. See ~\cite{figalli2021invitation} for more details on the Wasserstein distance. Given $\mu,  \nu \in \mathcal{P}_2(\mathbb{R})$,  the $2$-Wasserstein distance  between $\mu$ and $\nu$ is given as   
\begin{equation}
W_2(\mu,\nu) := \left(\underset{\gamma \in \Gamma(\mu,\nu)}{\operatorname{inf}}  \int_{\mathbb{R} \times \mathbb{R}}  \left \vert x-y\right \vert^2 \ d\gamma(x,y)\right)^{1/2},
\label{W2-distance}
\end{equation}
where $\Gamma(\mu,\nu)$ is the set of transport couplings with marginals $\mu$ and $\nu$ defined as
$$ \Gamma(\mu,\nu) :=  \left \{ \gamma \in \mathcal{P}(\mathbb{R} \times \mathbb{R}): {\pi_{{ x}}}_{\#} \gamma = \mu, \ {\pi_{{ y}}}_{\#} \gamma = \nu \right \},$$
and $\pi_x$ and $\pi_y$ are projections: for $(x,y) \in \mathbb{R}^2$, $\pi_x(x,y)=x$ and $\pi_y(x,y)=y$. 
Note that the $2$-Wasserstein distance on $\mathbb{R}$ has a quantile representation and is often used in numerical studies. Given $\mu, \nu \in  \mathcal{P}_2(\mathbb{R})$, let $F$ and $G$ denote the associated cumulative distribution functions:
\begin{equation*}
F(x)=\mu((-\infty,x]) \ \text{and} \ G(x)=\nu((-\infty,x]).
\end{equation*}
The generalized inverses of $F$ and $G$, also called the \emph{quantile functions}, are defined by
\begin{equation*}
F^{-1}(t)=\inf\left\{x\in\mathbb{R}:F(x)\geq t\right\},
\ \ 0<t<1,
\end{equation*}
and
\begin{equation*}
G^{-1}(t)=\inf\left\{x\in\mathbb{R}:G(x)\geq t\right\},
\ \ 0<t<1.
\end{equation*}
Then the 2-Wasserstein distance between $\mu$ and $\nu$ satisfies
\begin{equation}\label{eqn:W2_Quantile}
W_2(\mu,\nu)=\left(\int_0^1|F^{-1}(t)-G^{-1}(t)|^2 dt\right
)^{1/2}.
\end{equation}

In particular, given data $\{x_i\}_{i=1}^M$ and $\{y_i\}_{i=1}^N$ where $M$ may not equal $N$, let $\mu_M$ and $\nu_N$ be the corresponding empirical measures, i.e., $\mu_M:=\frac{1}{M}\sum\limits_{i=1}^M\delta_{x_i}$ and $
\nu_N:=\frac{1}{N}\sum\limits_{i=1}^N\delta_{y_i}$.
Then the associated quantile functions $F^{-1}$ and $G^{-1}$ are stepwise functions and have the form
\begin{align*}  
F^{-1}(t)
=\left\{\begin{array}{lr}x_{(1)}, & 0 < t \leq 1/M \\ x_{(2)}, & 1/M < t \leq 2/M \\ \vdots & \\ x_{(M)}, & (M-1)/M < t \leq 1 \end{array} \right.
 \ \text{and} \  
G^{-1}(t)=\left\{\begin{array}{lr}y_{(1)}, & 0 < t \leq 1/N \\ y_{(2)}, & 1/N < t \leq 2/N \\ \vdots & \\ y_{(N)}, & (N-1)/N < t \leq 1 \end{array} \right.
\end{align*}
where $
x_{(1)} \leq x_{(2)} \leq \cdots \leq x_{(M)}$ and $y_{(1)} \leq y_{(2)} \leq \cdots \leq y_{(N)}$ are order statistics.

Let $I_i=\left(\frac{i-1}{M},\frac{i}{M}\right]$ and $J_j=\left(\frac{j-1}{N},\frac{j}{N}\right]$,
for $i=1,\cdots,M$ and $j=1,\cdots,N$.
The length of the intersection $I_i\cap J_j$ is
\[
w_{ij}:=|I_i\cap J_j|
=
\left[
\min\left\{\frac{i}{M},\frac{j}{N}\right\}
-
\max\left\{\frac{i-1}{M},\frac{j-1}{N}\right\}
\right]_+,
\]
where $[x]_+:=\max\{x,0\}$. Therefore, Equation \ref{eqn:W2_Quantile} implies
\begin{equation}\label{eqn:sample_W2_unequal_size}
W_2^2(\mu_M,\nu_N)
=\sum_{i=1}^M\sum_{j=1}^N
\int_{I_i\cap J_j}
\left|x_{(i)}-y_{(j)}\right|^2 dt =\sum_{i=1}^M\sum_{j=1}^N
w_{ij}\left|x_{(i)}-y_{(j)}\right|^2.
\end{equation}
Especially, when $M=N$, the 2-Wasserstein distance between $\mu_N$ and $\nu_N$ satisfies
\begin{equation}\label{eqn:sample_W2}
W_2^2(\mu_N,\nu_N)
=\frac{1}{N}\sum_{i=1}^N
|x_{(i)}-y_{(i)}|^2.
\end{equation}

Similar to the decomposition of squared RMSE, the squared 2-Wasserstein distance admits the following decomposition
\begin{equation}
W_2^2(\mu,\nu)=(m_\mu-m_\nu)^2+W_2^2(\widetilde{\mu},\widetilde{\nu}),\label{equation:2W-Distance}
\end{equation}
where $\widetilde{\mu}$ and $\widetilde{\nu}$ are the centered measures of $\mu$ and $\nu$, respectively. 
Furthermore, \cite{irpino2007optimal} showed that the distribution differences in terms of squared 2-Wasserstein distance can be decomposed into three parts, including bias (or location), size (or variability), and shape. That is, 
\begin{equation}\label{eqn:w2_decompostition}
\begin{aligned}
W_2^2(\mu,\nu)
&=\underbrace{(m_\mu-m_\nu)^2}_{\text{bias}}+
\underbrace{(\sigma_\mu-\sigma_\nu)^2}_{\text{size}}+
\underbrace{2\sigma_\mu\sigma_\nu(1-\rho_q)}_{\text{shape}},
\end{aligned}
\end{equation}
where $\rho_q$ is the \emph{quantile correlation coefficient} between $\mu$ and $\nu$ defined by
\begin{equation*}
\rho_q=\frac{\displaystyle\int_0^1\bigl(F^{-1}(t)-m_\mu\bigr)\bigl(G^{-1}(t)-m_\nu\bigr)\,dt}{\sigma_\mu\sigma_\nu}.
\end{equation*}
 See \cite{chen2026taylor, irpino2015basic,irpino2015linear} for more applications of such decomposition.

\section{Wasserstein Target Diagram}\label{sec:wtarget}

Given predictions  $\{x_i\}_{i=1}^N$ and observations $\{y_i\}_{i=1}^N$, suppose $\bar{x}$ and $\bar{y}$ are corresponding sample means, $\sigma_x$ and $\sigma_y$ are associated sample standard deviations, and $\{x_i'=x_i-\bar{x}\}_{i=1}^N$ and $\{y_i'=y_i-\bar{y}\}_{i=1}^N$ are their centered data. Then we have the following decomposition 
\begin{equation*}
\operatorname{RMSE}^2=\operatorname{Bias}^2 + \operatorname{CRMSE}^2, 
\end{equation*}
where $\operatorname{Bias}=\bar{x}-\bar{y}$, CRMSE is the centered RMSE, with
\begin{equation*}
\begin{aligned}
\operatorname{RMSE}^2
=\frac{1}{N}\sum_{i=1}^N(x_i-y_i)^2  \ \text{and} \ 
\operatorname{CRMSE}^2
=\frac{1}{N}\sum_{i=1}^N(x_i'-y_i')^2.
\end{aligned}
\end{equation*}
Then the point in the target diagram for
$\{x_i\}_{i=1}^N$ and $\{y_i\}_{i=1}^N$ has coordinates $(s, t)$ where
\begin{equation*}
s=\text{sign}(\sigma_x-\sigma_y) \cdot \operatorname{CRMSE} \ \text{and} \
t=\operatorname{Bias},
\end{equation*}
so that when $\sigma_x\neq\sigma_y$, $\text{RMSE}^2 = s^2+t^2$. Therefore, the radial distance from the point $(s,t)$ to the origin in the target diagram represents the associated RMSE.

The target diagram requires paired observed and predicted data of equal length, and cannot be directly applied when the two samples have different sizes. For example, only observed data exist in the wind speed distribution fitting \cite{yang2025beta}, and the target diagram does not work.  In contrast, the Wasserstein distance compares probability distributions directly and does not require paired data or equal sample sizes.
This motivates us to lift the target diagram into the probability level via the Wasserstein distance and construct an analogous Wasserstein target diagram. The mathematical details are as follows. 

Let $\mu,\nu\in\mathcal{P}_2(\mathbb{R})$ with means
$m_\mu,m_\nu$ and standard deviations $\sigma_\mu,\sigma_\nu$,
respectively. Let $\widetilde{\mu}$ and $\widetilde{\nu}$ denote their
centered measures. Equation~\ref{equation:2W-Distance} shows that
\begin{equation}
W_2^2(\mu,\nu)
=(m_\mu-m_\nu)^2
+W_2^2(\widetilde{\mu},\widetilde{\nu}).
\end{equation}
Then the coordinates of a point $(s, t)$ in the Wasserstein target diagram with respect to $\mu$ and $\nu$ are given as
\begin{equation}
s=\text{sign}(\sigma_\mu-\sigma_\nu)\cdot
W_2(\widetilde{\mu},\widetilde{\nu}) \ \text{and}
\ t=m_\mu-m_\nu,
\end{equation}
where $s$ is the signed centered 2-Wasserstein distance ($W_2$) between $\mu$ and $\nu$, and $t$ is the bias. When $\sigma_\mu \neq \sigma_\nu$, the coordinates satisfy
\begin{equation}
\begin{aligned}
s^2+t^2
&=W_2^2(\widetilde{\mu},\widetilde{\nu})
+(m_\mu-m_\nu)^2=W_2^2(\mu,\nu),
\end{aligned}
\end{equation}
which shows that the radial distance from point $(s,t)$ to the origin in the Wasserstein target diagram represents the 2-Wasserstein distance $W_2(\mu, \nu)$.

\begin{figure*}[ht]
\centering
\includegraphics[width=1\textwidth]{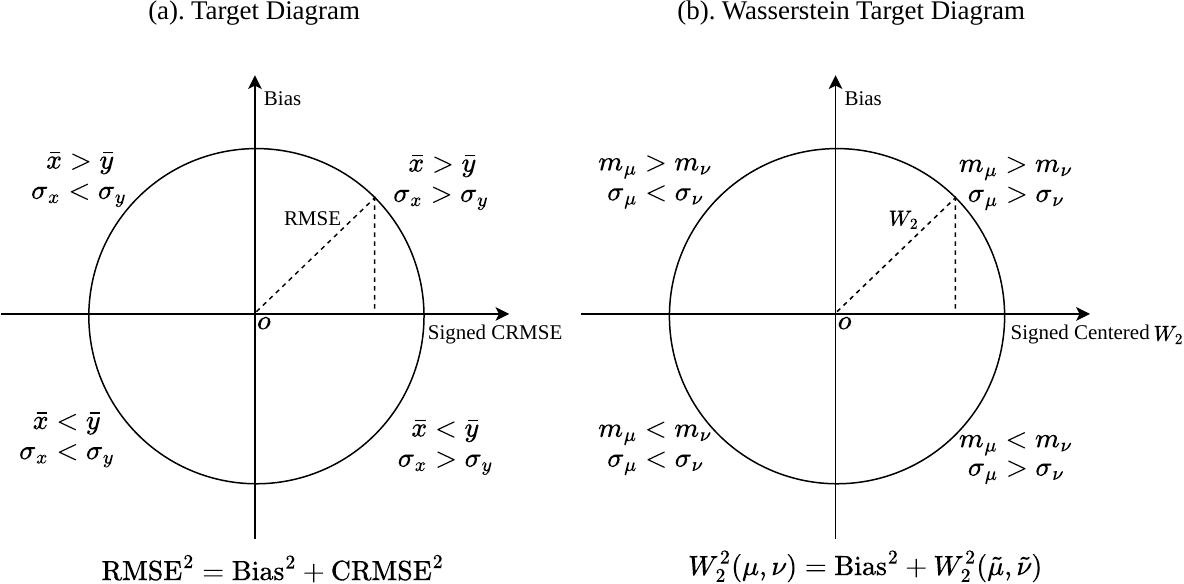}
\caption{Comparison of the Target Diagram and Wasserstein Target Diagram. The target diagram evaluates models based on the temporal or spatial matching between predicted and observed data, while the Wasserstein target diagram is from the probability distribution perspective. Panel (a): Given data $\{x_i\}_{i=1}^N$ and $\{y_i\}_{i=1}^N$ with sample means $\bar{x}$ and $\bar{y}$ and different standard deviations $\sigma_x$ and $\sigma_y$, the target diagram is constructed by the decomposition $\operatorname{RMSE}^2
=\operatorname{Bias}^2+\operatorname{CRMSE}^2$, where the horizontal and vertical coordinates are $\operatorname{sign}(\sigma_x-\sigma_y)\cdot \operatorname{CRMSE}$ and the bias $\bar{x}-\bar{y}$, respectively, where CRMSE is the centered RMSE. The radial distance to the origin is RMSE. Panel (b):  Given probability distributions $\mu, \nu \in \mathcal{P}_2(\mathbb{R})$ with means $m_\mu$ and $m_\nu$ and different standard deviations $\sigma_\mu$ and $\sigma_\nu$, the Wasserstein target diagram is constructed by the decomposition $W_2^2(\mu, \nu) = \operatorname{Bias}^2 + W_2^2(\Tilde{\mu}, \Tilde{\nu})$, where the horizontal and vertical coordinates are $\operatorname{sign}(\sigma_\mu-\sigma_\nu)\cdot  W_2(\Tilde{\mu}, \Tilde{\nu})$ and the bias $m_\mu-m_\nu$, respectively. The radial distance is the 2-Wasserstein distance $W_2(\mu, \nu)$.}
\label{fig:target_compare}
\end{figure*}

Both the target diagram and Wasserstein target diagram use the bias as the vertical coordinate. For the horizontal coordinate and radial distance to the origin, the target diagram uses signed CRMSE and RMSE, while the Wasserstein target diagram uses signed centered 2-Wasserstein distance and 2-Wasserstein distance. 
Note that in target and Wasserstein target diagrams, the sign of horizontal coordinate indicates whether the predictions have a larger or smaller standard deviation than the observations, while the sign of the vertical coordinate indicates whether the predictions have a larger or smaller mean than the observations.
In particular, the magnitude of the horizontal coordinate in the Wasserstein target diagram reflects discrepancies between the centered distributions, including distribution differences in size and shape (see Equations~\ref{equation:2W-Distance} and~\ref{eqn:w2_decompostition}). Figure \ref{fig:target_compare} shows the comparison of target diagram and Wasserstein target diagram.

When evaluating models, the target diagram uses paired data of equal length and shows the spatial or temporal matching between predictions and observations, while the Wasserstein target diagram uses the 2-Wasserstein distance and compares the distribution similarities of used models to the reference. Therefore, the Wasserstein target diagram allows multiple probability distributions to be compared with a reference within a common coordinate system, and the closer a distribution is to the origin in the diagram, the smaller its 2-Wasserstein distance and the better its performance. Unlike the target diagram, the Wasserstein target diagram does not require paired observations or equal sample size and can be directly applied to evaluate the probability distribution fitting of given data. Furthermore, Theorem \ref{thm:coincidence} shows that the target diagram and the Wasserstein target diagram coincide if and only if
the given data $\{x_i\}_{i=1}^N$ and $\{y_i\}_{i=1}^N$ are comonotone, where the data are taken as equally weighted empirical measures in the Wasserstein target diagram. The proof is based on the rearrangement inequality. 
Recall that given $\{x_i\}_{i=1}^N$ and $\{y_i\}_{i=1}^N$ with order statistics $\{x_{(i)}\}_{i=1}^N$ and $\{y_{(i)}\}_{i=1}^N$, that is, 
$$
x_{(1)} \leq x_{(2)} \leq \cdots \leq x_{(N)} \ \text{and} \ y_{(1)} \leq y_{(2)} \leq \cdots \leq y_{(N)},
$$
the rearrangement inequality claims  
\begin{equation*}
\sum_{i=1}^N x_i y_i\leq\sum_{i=1}^N x_{(i)}y_{(i)}, 
\end{equation*}
where the equality holds if and only if $\{x_i\}_{i=1}^N$ and $\{y_i\}_{i=1}^N$ are comonotone, i.e.,
\begin{equation*}
(x_i-x_j)(y_i-y_j) \geq 0, \ \text{for any} \ i,j. 
\end{equation*}
Then we have the following main result in this paper. 

\begin{theorem}\label{thm:coincidence}
Given data $\{x_i\}_{i=1}^N$ and $\{y_i\}_{i=1}^N$ with different standard deviations, let
\begin{equation*}
\mu_N:=\frac{1}{N}\sum_{i=1}^N\delta_{x_i} \ \text{and} \ 
\nu_N:=\frac{1}{N}\sum_{i=1}^N\delta_{y_i}.
\end{equation*}
The Wasserstein target diagram for $\mu_N$ and $\nu_N$ coincides with the target diagram for $\{x_i\}_{i=1}^N$ and $\{y_i\}_{i=1}^N$ if and only if  $\{x_i\}_{i=1}^N$ and $\{y_i\}_{i=1}^N$ are comonotone, i.e.,
\begin{equation*}
(x_i-x_j)(y_i-y_j) \geq 0, \ \text{for any} \ i,j. 
\end{equation*}
\end{theorem}

\begin{proof}
 By Equation \ref{eqn:sample_W2}, the 2-Wasserstein distance between $\mu_N$ and $\nu_N$ satisfies 
\begin{equation*}
W_2^2(\mu_N,\nu_N)
=\frac{1}{N}\sum_{i=1}^N
|x_{(i)}-y_{(i)}|^2=\frac{1}{N}\sum_{i=1}^N x_i^2
+\frac{1}{N}\sum_{i=1}^N y_i^2-\frac{2}{N}
\sum_{i=1}^N x_{(i)}y_{(i)},
\end{equation*}
where $
x_{(1)} \leq x_{(2)} \leq \cdots \leq x_{(N)}$ and $ y_{(1)} \leq y_{(2)} \leq \cdots \leq y_{(N)}$.
Then by the rearrangement inequality, 
\begin{equation*}
\begin{aligned}
W_2^2(\mu_N,\nu_N)
&\leq
\frac{1}{N}\sum_{i=1}^N x_i^2
+\frac{1}{N}\sum_{i=1}^N y_i^2-\frac{2}{N}\sum_{i=1}^N x_i y_i\\
&=\frac{1}{N}\sum_{i=1}^N
(x_i-y_i)^2 = \operatorname{RMSE}^2,
\end{aligned}
\end{equation*}
and the equality holds if and only if
$\{x_i\}_{i=1}^N$ and $\{y_i\}_{i=1}^N$ are comonotone, i.e.,
\begin{equation*}
(x_i-x_j)(y_i-y_j) \geq 0, \ \text{for any} \ i,j. 
\end{equation*}

Suppose $\bar{x}$ and $\bar{y}$ are the means of $\{x_i\}_{i=1}^N$ and $\{y_i\}_{i=1}^N$, and 
$\{x_i' = x_i-\bar{x}\}_{i=1}^N$ and $\{y_i' = y_i-\bar{y}\}_{i=1}^N$ are the associated centered data. Let $\widetilde{\mu}_N$ and $\widetilde{\nu}_N$ denote the
centered measures of $\mu_N$ and $\nu_N$, i.e., 
\begin{equation*}
\widetilde{\mu}_N:=\frac{1}{N}\sum_{i=1}^N\delta_{x'_i} \ \text{and} \ 
\widetilde{\nu}_N:=\frac{1}{N}\sum_{i=1}^N\delta_{y'_i}.
\end{equation*}
Then applying the same argument to $\widetilde{\mu}_N$ and $\widetilde{\nu}_N$ gives
\begin{equation*}
W_2^2(\widetilde{\mu}_N,\widetilde{\nu}_N)
\leq \frac{1}{N}\sum_{i=1}^N
(x_i'-y_i')^2 = 
\operatorname{CRMSE}^2,
\end{equation*}
and the equality holds if and only if
\begin{equation*}
(x_i'-x_j')(y_i'-y_j') = (x_i-x_j)(y_i-y_j) \geq 0,
 \ \text{for any} \ i,j,
\end{equation*}
which means the equality holds if and only if
$\{x_i\}_{i=1}^N$ and $\{y_i\}_{i=1}^N$ are comonotone.
Therefore, under the comonotone condition,
\begin{equation*}
W_2(\mu_N,\nu_N) = \operatorname{RMSE} \ \text{and} \
W_2(\widetilde{\mu}_N,\widetilde{\nu}_N)
=\operatorname{CRMSE}.
\end{equation*}
Moreover,
\begin{equation*}
m_{\mu_N}=\bar{x}, m_{\nu_N} = \bar{y}, \sigma_{\mu_N} = \sigma_x, \ \text{and} \ \sigma_{\nu_N}= \sigma_y,
\end{equation*}
where $\sigma_x$ and $\sigma_y$ are sample standard deviations of $\{x_i\}_{i=1}^N$ and $\{y_i\}_{i=1}^N$, respectively.
Then the target diagram and Wasserstein target diagram have the same
vertical coordinate, horizontal magnitude, and horizontal sign. Hence, they
coincide. 

Conversely, if the target diagram and Wasserstein target diagram coincide, then the corresponding horizontal coordinates are the same, i.e., 
\begin{equation*}
    \text{sign}(\sigma_x-\sigma_y) \cdot \operatorname{CRMSE} = \text{sign}(\sigma_{\mu_N}-\sigma_{\nu_N}) \cdot W_2(\widetilde{\mu}_N,\widetilde{\nu}_N).
\end{equation*}
Note that $\sigma_{\mu_N} = \sigma_x $, $ \sigma_{\nu_N}= \sigma_y$, and $\sigma_x \neq \sigma_y$. Then $\text{sign}(\sigma_x-\sigma_y) =  \text{sign}(\sigma_{\mu_N}-\sigma_{\nu_N}) = \pm1$ and thus
$W_2(\widetilde{\mu}_N,\widetilde{\nu}_N)
=\operatorname{CRMSE}$,
which implies the equality condition in the rearrangement inequality:
\begin{equation*}
(x_i-x_j)(y_i-y_j) \geq 0,
 \ \text{for any} \ i,j.
\end{equation*}
\end{proof}

In climate science, the cumulative climate variables indexed by time are nondecreasing and provide natural examples of comonotone structures, such as cumulative precipitation, accumulated runoff, and cumulative solar radiation. 
Therefore, for such cumulative variables, the Wasserstein target diagram just becomes the original target diagram. Note that when the given data $\{x_i\}_{i=1}^N$ and $\{y_i\}_{i=1}^N$ have the same sample standard deviations, i.e., $\sigma_x=\sigma_y$, the horizontal coordinate is zero by the definition of the sign function, regardless of the value of CRMSE or $W_2(\widetilde{\mu},\widetilde{\nu})$. Consequently, the point on the vertical axis may represent equal sample standard deviations,  zero CRMSE, or zero $W_2(\widetilde{\mu},\widetilde{\nu})$, and the radial distance does not necessarily equal RMSE or $W_2(\mu, \nu)$. This special case should be particularly considered when it appears in both target and Wasserstein target diagrams.

\section{Applications of the Wasserstein Target Diagram}\label{sec:simulation}

In this section, we illustrate the Wasserstein target diagram through two simulations and a wind speed application. The first simulation compares the target and Wasserstein target diagrams using observed and predicted samples of equal size. The second simulation considers unequal sample sizes, for which the original target diagram does not apply directly. Finally, we apply the Wasserstein target diagram to wind speed data to compare fitted probability distributions with observed data. 
Throughout this section, $N$ denotes a normal distribution parameterized by mean and variance, $LogN$ denotes a lognormal distribution parameterized by log-scale mean and variance, and $GAM$ and $W$ denote gamma and Weibull distributions parameterized by shape and scale, respectively. The parameters of these distributions in the simulation studies are also compared with the reference distribution. In addition, we use $W_2$ for the 2-Wasserstein distance. 

\subsection{Simulation with Equal Sample Sizes: Comparison of Target Diagrams}

To illustrate the Wasserstein target diagram and compare it with the original target diagram, we generate an observed sample of size $n=1{,}000$ from a normal distribution $N(4, 4)$, with theoretical mean 4 and standard deviation 2. The observed sample has a mean of 3.920872 and standard deviation of 2.001575. We independently generate nine prediction samples from the following distributions, each with the same sample size $n=1{,}000$:

\begin{itemize}
    \item $N(4,4)$: the same underlying normal distribution as the observed sample.
    \item $N(3,2.25)$: smaller mean $3$ and standard deviation $1.5$.
    \item $N(5,6.25)$: larger mean $5$ and standard deviation $2.5$.
    \item $N(5,2.25)$: larger mean $5$ but smaller standard deviation $1.5$.
    \item $N(3,6.25)$: smaller mean $3$ but larger standard deviation $2.5$.
    \item $GAM(5,1)$: a gamma distribution with shape parameter $5$ and scale parameter $1$, having larger mean $5$ and larger standard deviation $\sqrt{5}\approx2.236$.
    \item $W(6,4)$: a Weibull distribution with shape parameter $6$ and scale parameter $4$, having smaller mean $\approx3.711$ and standard deviation $\approx0.719$.
    \item $LogN(1.5,0.2025)$: a lognormal distribution with log-scale mean $1.5$ and log-scale standard deviation $0.45$, having larger mean $\approx4.959$ and standard deviation $\approx2.350$.
    \item $Bimodal$: an equally weighted mixture $\frac{1}{2}N(3,0.25)+\frac{1}{2}N(6,0.25)$, having larger mean $4.5$ but smaller standard deviation $\sqrt{2.5}\approx1.581$.
\end{itemize}

\begin{table*}[h]
\centering
\caption{Wasserstein Target Diagram Measures for Samples of Equal Size. The reference sample consists of $n=1{,}000$ observations from the normal distribution $N(4,4)$, with a sample mean 3.920872 and standard deviation 2.001575. The predicted sample generated from each distribution also contains $1{,}000$ data points.}
\label{table:wTargetSimulation_equalLength}
\begin{tabular}{lrrrr}
\toprule
Models & $n$ & Signed centered $W_2$ & Bias & $W_2$ \\
\midrule
$N(4,4)$      & 1,000 & -0.144483 &  0.095907 & 0.173417 \\
$N(3,2.25)$   & 1,000 & -0.537894 & -0.905342 & 1.053078 \\
$N(5,6.25)$   & 1,000 &  0.492475 &  1.245325 & 1.339166 \\
$N(5,2.25)$   & 1,000 & -0.506116 &  1.168765 & 1.273642 \\
$N(3,6.25)$   & 1,000 &  0.463521 & -0.967032 & 1.072382 \\
$GAM(5,1)$     & 1,000 &  0.468985 &  1.073327 & 1.171315 \\
$W(6,4)$       & 1,000 & -1.314335 & -0.222236 & 1.332991 \\
$LogN(1.5,0.2025)$    & 1,000 &  0.788539 &  1.125171 & 1.373974 \\
$Bimodal$   & 1,000 & -0.763342 &  0.590595 & 0.965139 \\
\bottomrule
\end{tabular}
\end{table*}

\begin{figure*}[h]
\centering
\includegraphics[scale=0.568]{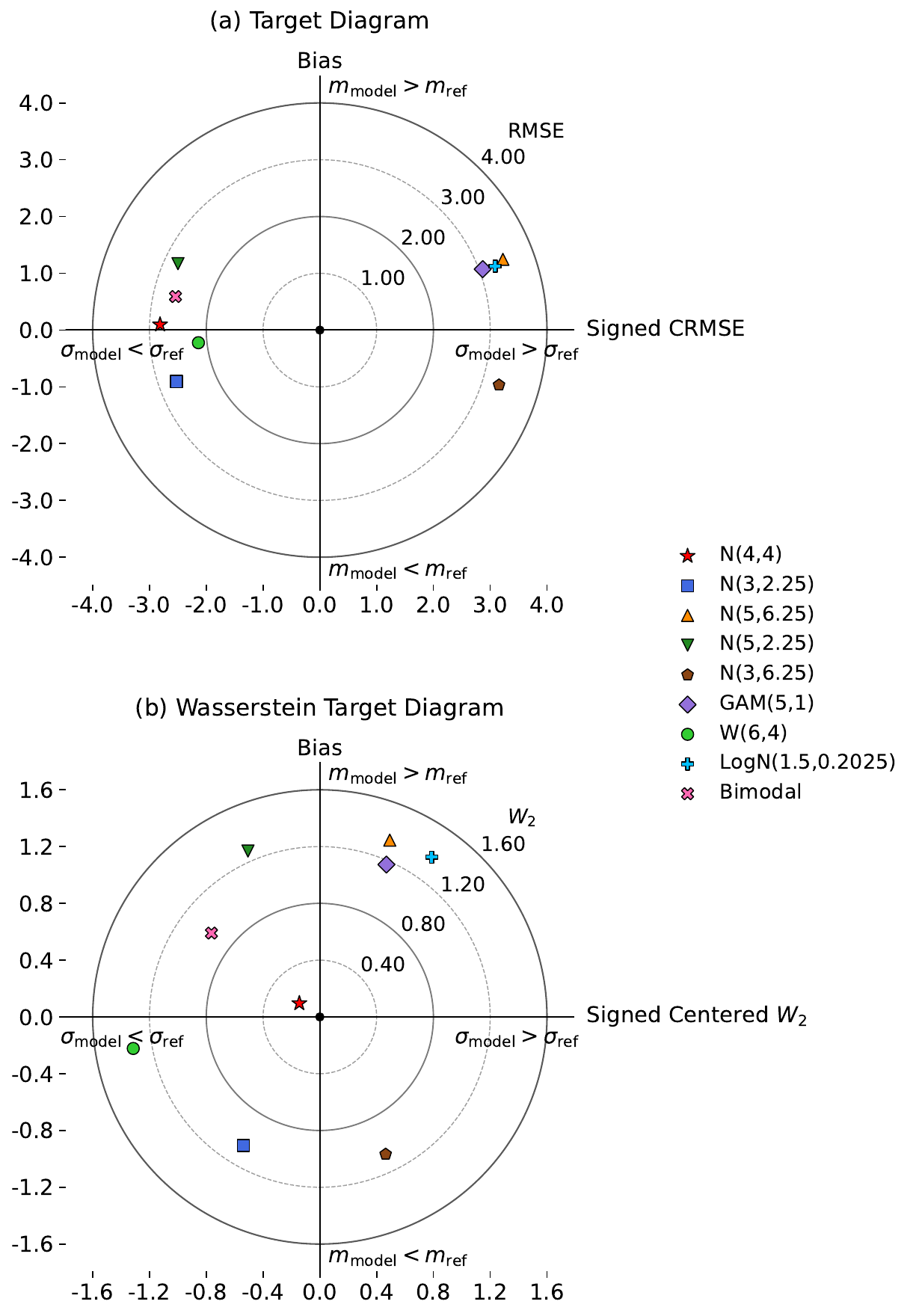}
\caption{Target Diagram and Wasserstein Target Diagram for Samples of Equal Size. The reference consists of $1{,}000$ observations from the normal distribution $N(4,4)$, and other predicted samples have the same size. The $N(4,4)$ predicted sample lies closest to the origin in the Wasserstein target diagram, but not in the target diagram. This shows the usefulness of Wasserstein target diagram for probability distribution comparison, even for samples from the same distribution.}
\label{fig:wTargetSimulation_equalLength}
\end{figure*}

As the sample generation suggests, the $N(4,4)$ predicted samples are generated from the same distribution as the observed samples, which means that the corresponding 2-Wasserstein distance ($W_2$) is small, but their RMSE may remain large because it depends on the pointwise pairing. In this case, the Wasserstein target diagram can detect the distributional similarities between predicted and observed data, but the target diagram cannot, as shown in Table~\ref{table:wTargetSimulation_equalLength}  and Figure~\ref{fig:wTargetSimulation_equalLength}.
Table~\ref{table:wTargetSimulation_equalLength} shows the biases, signed centered 2-Wasserstein distances, and 2-Wasserstein distances between predicted and observed data, where the 2-Wasserstein distance is computed by Equation \ref{eqn:sample_W2}. Figure~\ref{fig:wTargetSimulation_equalLength} further shows the associated target diagram and Wasserstein target diagram.  It can be seen from the Wasserstein target diagram that the $N(4,4)$ predicted samples are closest to the origin, with a 2-Wasserstein  distance of 0.173417. 
In contrast, the target diagram does not identify the $N(4,4)$ prediction as the closest to the origin, since the $N(4,4)$ prediction has larger RMSE and CRMSE with values of 2.815418 and 2.813784, respectively. This occurs because RMSE and CRMSE depend on the pointwise pairing of data but two independently generated samples have no natural one-to-one correspondence. 
Consequently, the target diagram does not identify that the $N(4,4)$ prediction has the highest similarity with the observed data, while the Wasserstein target diagram shows such identification. The remaining predictions in the Wasserstein target diagram exhibit larger distances to the origin due to differences in their biases, sizes, and shapes. For example, the Weibull prediction has a relatively small bias of -0.222236 but a large centered 2-Wasserstein  distance of 1.314335, indicating that bias alone does not adequately characterize its distributional discrepancy with the observed data. These results illustrate the usefulness of the Wasserstein target diagram for comparing probability distributions, even when samples are independently generated from the same distribution.

\subsection{Simulation with Unequal Sample Sizes: Distribution Comparison}

To illustrate the application of Wasserstein target diagram to samples of unequal sizes, we generate a reference sample of size $n=10{,}000$ from a gamma distribution $GAM(4,1)$ as the observed data. The reference distribution has theoretical mean $4$ and standard deviation $2$, while the generated sample has a mean $4.024726$ and standard deviation $1.997160$. We independently generate nine other samples from the following distributions as predicted data:

\begin{itemize}
    \item $GAM(4,1)$: $n=20{,}000$, the same underlying gamma distribution as the observed sample.
    \item $GAM(4,0.8)$: $n=15{,}000$, smaller mean $3.2$ and standard deviation $1.6$.
    \item $GAM(4,1.2)$: $n=25{,}000$, larger mean $4.8$ and standard deviation $2.4$.
    \item $GAM(9,0.5)$: $n=18{,}000$, larger mean $4.5$ but smaller standard deviation $1.5$.
    \item $GAM(4,1)+1$: $n=22{,}000$, a gamma distribution shifted by $1$, having larger mean $5$ but the same standard deviation $2$.
    \item $W(2,4.5)$: $n=30{,}000$, a Weibull distribution with shape parameter $2$ and scale parameter $4.5$, having a slightly smaller mean $\approx3.988$ but a slightly larger standard deviation $\approx2.085$.
    \item $LogN(1.285,0.2025)$: $n=25{,}000$, a lognormal distribution with log-scale mean $1.285$ and log-scale standard deviation $0.45$, having mean $\approx4$ and smaller standard deviation $\approx1.895$.
    \item $N(4,4)$: $n=35{,}000$, a normal distribution with the same theoretical mean $4$ and standard deviation $2$ as the reference distribution.
    \item $Bimodal$: $n=40{,}000$, an equally weighted mixture $\frac{1}{2}N(2.5,0.25)+\frac{1}{2}N(5.5,0.25)$, having the same mean $4$ but smaller standard deviation $\sqrt{2.5}\approx1.581$.
\end{itemize}

\begin{figure*}[h]
\centering
\includegraphics[scale=0.6]{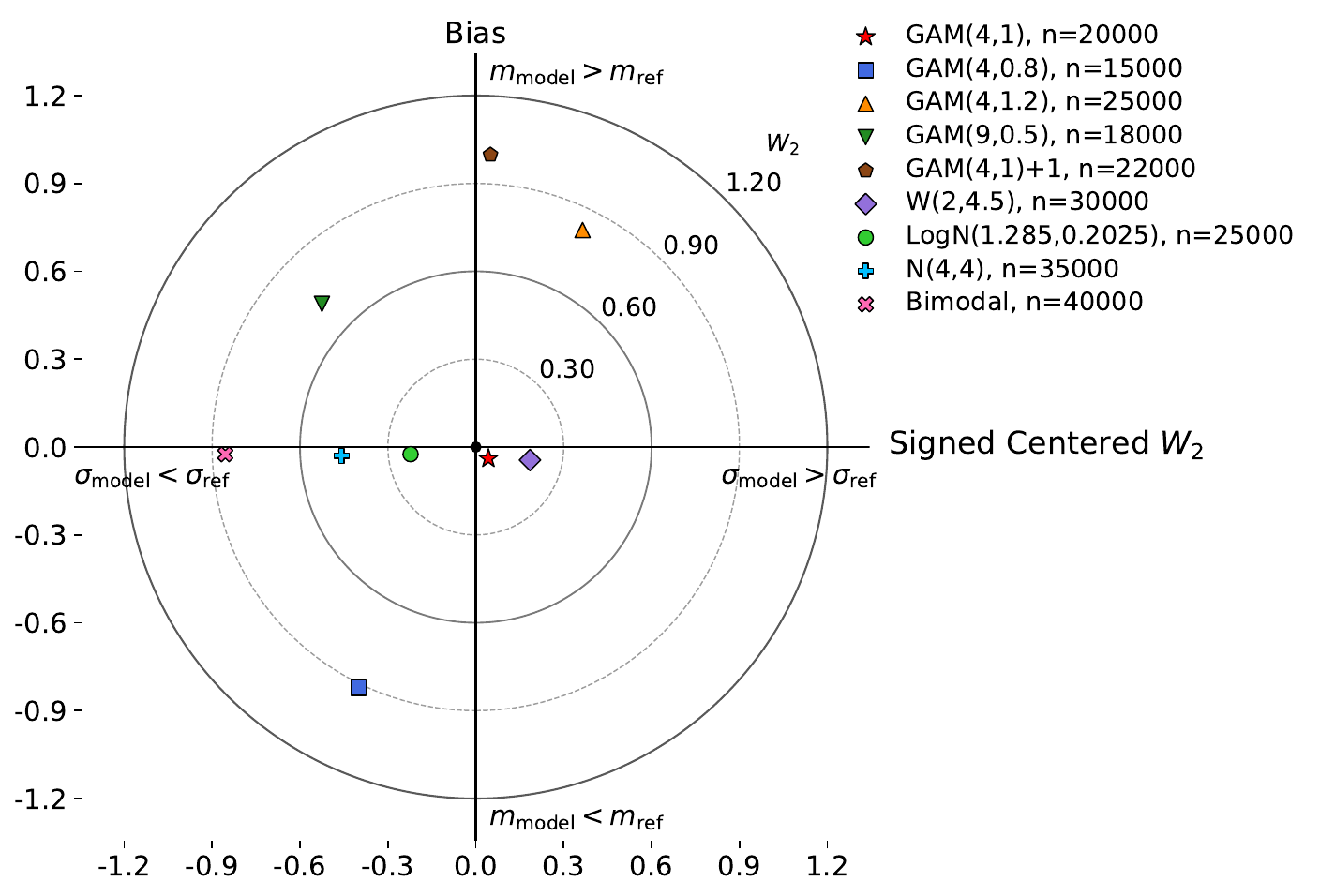}
\caption{Wasserstein Target Diagram for Samples of Different Sizes. The reference sample consists of $n=10{,}000$ observations from $GAM(4,1)$, with a mean $4.024726$ and standard deviation $1.997160$. The horizontal coordinate represents the signed centered 2-Wasserstein distance (signed centered $W_2$), the vertical coordinate represents the bias, and the radial distance represents the 2-Wasserstein distance ($W_2$). The $GAM(4,1)$ prediction with $n=20{,}000$ has the highest distribution similarity with the reference since it is closest to the origin.}
\label{fig:wTargetSimulation}
\end{figure*}

\begin{table*}[h]
\centering
\caption{Wasserstein Target Diagram Measures for Samples of Unequal Sizes. The reference sample consists of
$n=10{,}000$ observations from the Gamma distribution $GAM(4,1)$, with a sample mean $4.024726$ and standard deviation $1.997160$. Predicted sample sizes are listed in the second column.}
\label{table:wTargetSimulation}
\begin{tabular}{lrrrr}
\toprule
Prediction distribution & Size $n$ & Signed centered $W_2$ & Bias & $W_2$ \\
\midrule
$\mathrm{GAM}(4,1)$          & 20,000 &  0.043151 & -0.039307 & 0.058370 \\
$\mathrm{GAM}(4,0.8)$        & 15,000 & -0.400569 & -0.821487 & 0.913945 \\
$\mathrm{GAM}(4,1.2)$        & 25,000 &  0.364501 &  0.740879 & 0.825689 \\
$\mathrm{GAM}(9,0.5)$        & 18,000 & -0.525069 &  0.488746 & 0.717336 \\
$\mathrm{GAM}(4,1)+1$        & 22,000 &  0.050082 &  0.997958 & 0.999214 \\
$\mathrm{W}(2,4.5)$          & 30,000 &  0.185145 & -0.044270 & 0.190364 \\
$\mathrm{LogN}(1.285,0.2025)$ & 25,000 & -0.222552 & -0.024330 & 0.223878 \\
$N(4,4)$                    & 35,000 & -0.457811 & -0.029927 & 0.458788 \\
$\text{Bimodal}$             & 40,000 & -0.854919 & -0.024787 & 0.855278 \\
\bottomrule
\end{tabular}
\end{table*}

Table~\ref{table:wTargetSimulation} summarizes the resulting measures, where the 2-Wasserstein distance is calculated by Equation \ref{eqn:sample_W2_unequal_size}, and Figure~\ref{fig:wTargetSimulation} presents the corresponding Wasserstein target diagram. Note that
the prediction generated from $GAM(4,1)$ is closest to the origin in the diagram, with a 2-Wasserstein distance of 0.058370, despite the sample size being doubled compared to the observation. This is because the predicted and observed data follow the same distribution. The $GAM(4,0.8)$ prediction appears in the lower-left region because of its smaller mean and standard deviation, whereas $GAM(4,1.2)$ appears in the upper-right region. The $GAM(9,0.5)$ prediction combines a larger mean with smaller variability and appears in the upper-left region. The shifted Gamma prediction has a bias of 0.997958  but a centered 2-Wasserstein distance of only 0.050082, indicating a discrepancy primarily associated with bias.

The Weibull, lognormal, normal, and bimodal predictions have small absolute biases but different centered 2-Wasserstein distances. In particular, the normal distribution has the same theoretical mean and standard deviation as the reference, but its centered 2-Wasserstein distance is 0.457811, demonstrating its sensitivity to shape beyond the first two moments. Its negative horizontal coordinate reflects a standard deviation slightly below that of the reference sample. The bimodal distribution also has theoretical mean 4, but its centered 2-Wasserstein distance of 0.854919
reflects distribution differences in both size and shape.

\subsection{Wind Speed Data: Comparison of Fitted Distributions}\label{sec:wind}

Accurate modeling of wind speed distributions is important for a variety of applications, including building design and wind energy assessment  \cite{yang2025beta}. A key task is to compare which candidate probability distribution can accurately estimate the wind speed distribution, and the Wasserstein target diagram provides a direct assessment for such distribution comparison.  In this section, we use processed wind speed data described in \cite{yang2025beta}, which are hourly averaged wind speed measurements collected at a height of 80 meters (m) from the year 2010 to 2020 at the Flatirons M2 meteorological tower, located about 8 kilometers south of Boulder, Colorado, USA \cite{jager1996nrel, M2data}. The observed sample contains $n = 96,432$ measurements, with a mean $4.638$ m/s and a standard deviation of $3.474$ m/s. 

We compare ten fitted distributions using the maximum likelihood estimates reported in Table~4 of \cite{yang2025beta}, including beta-generalized Lindley (BGL), beta-exponential (BE), beta-Weibull (BW), beta-Lindley (BL), generalized Lindley (GL), Lindley (L), lognormal (LogN), gamma (GAM), Weibull (W), and generalized extreme value (GEV) distributions. Their notation and probability density functions appear in Table~3 of the same reference \cite{yang2025beta}. To find the 2-Wasserstein distance between each fitted distribution and the empirical distribution of wind speed, we first generate a sample of the same size $n = 96,432$ from each fitted distribution and then use Equation \ref{eqn:sample_W2} to calculate the corresponding 2-Wasserstein distance.

\begin{figure}[ht]
    \centering
    \includegraphics[scale=0.6]{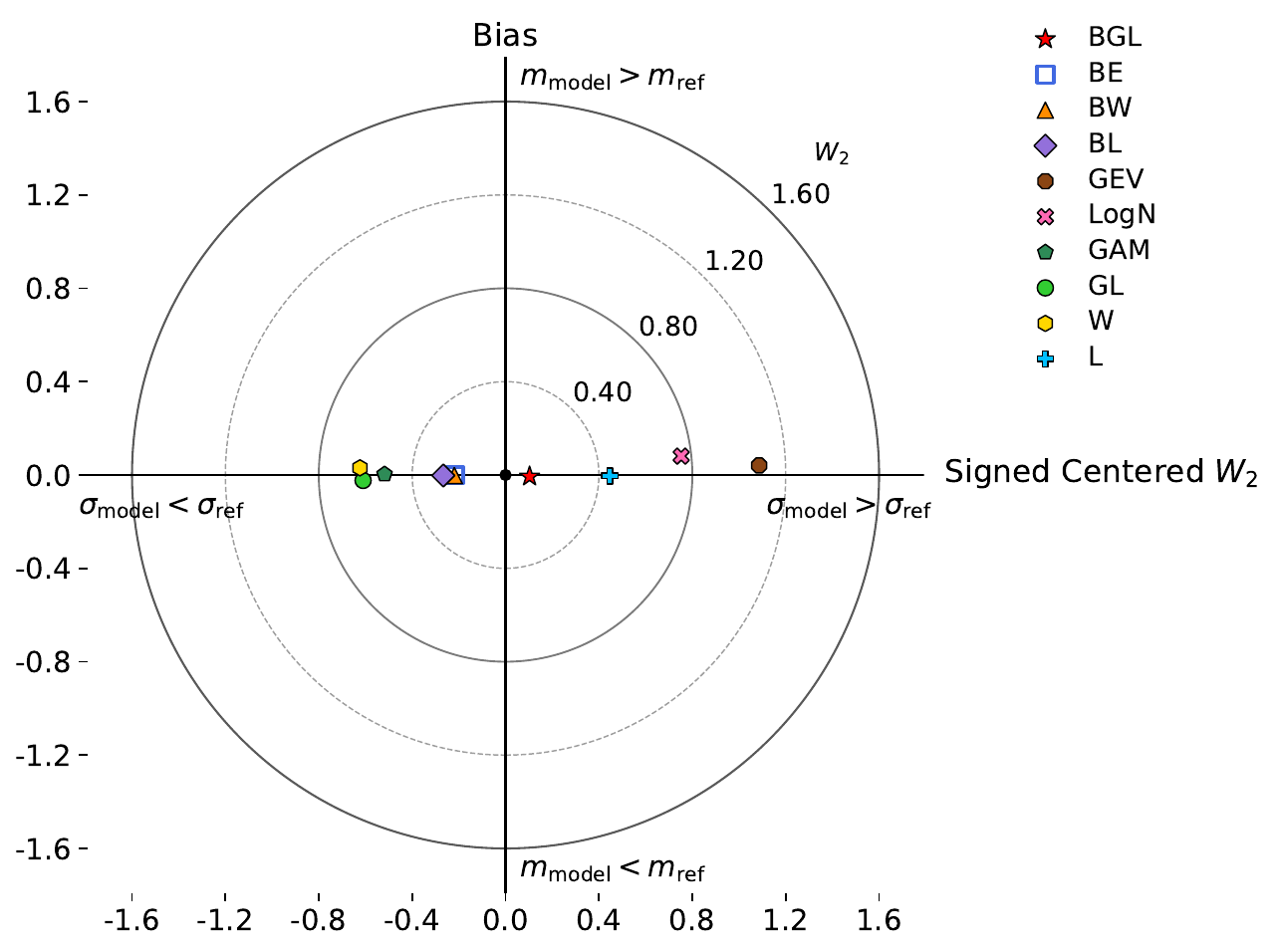} 
    \caption{Wasserstein Target Diagram for Distribution Fitting of Wind Speed. The reference wind speed sample has mean $4.638$ and standard deviation $3.474$. The horizontal coordinate represents the signed centered 2-Wasserstein distance (signed centered $W_2$), the vertical coordinate represents the bias, and the radial distance represents the 2-Wasserstein distance ($W_2$). Ten distributions are compared and BGL has the best performance, since it is closest to the origin and has the smallest 2-Wasserstein distance. } 
    \label{fig:wTargetWind}
\end{figure}

\begin{table*}[ht]
\centering
\caption{Wasserstein Target Diagram Measures for Distribution Fitting of Wind Speed. The observed wind speed sample has a mean of $4.638$ and a standard deviation of $3.474$. The fitted parameters of candidate distributions are the maximum likelihood estimates in Table~4 of \cite{yang2025beta}.}
\label{table:wTargetWind}
\begin{tabular}{lccc}
\toprule
Model & Signed centered $W_2$ & Bias & $W_2$ \\
\midrule
$BGL$ &  0.102954 & -0.006219 & 0.103142 \\
$BE$        & -0.216582 & -0.000999 & 0.216585 \\
$BW$  & -0.219444 & -0.002562 & 0.219459 \\
$BL$        & -0.266436 & -0.002081 & 0.266444 \\
$GL$             & -0.610161 & -0.023248 & 0.610604 \\
$L$                   &  0.446706 & -0.003918 & 0.446723 \\
$LogN$          &  0.752523 &  0.080740 & 0.756842 \\
$GAM$            & -0.518327 &  0.003346 & 0.518338 \\
$W$              & -0.623616 &  0.030635 & 0.624368 \\
$GEV$       &  1.086230 &  0.040883 & 1.086999 \\
\bottomrule
\end{tabular}
\end{table*}

Table~\ref{table:wTargetWind} summarizes the signed centered 2-Wasserstein distance, bias, and 2-Wasserstein distance for each probability distribution, and Figure~\ref{fig:wTargetWind} presents the corresponding Wasserstein target diagram. In the diagram, all the models are close to the horizontal axis, which means that the fitted distributions generally have small biases. Therefore, absolute bias alone provides limited information to evaluate the candidate distributions. For example, BE has a smaller absolute bias than BGL ($0.000999$ versus $0.006219$), but BGL has a smaller centered 2-Wasserstein distance ($0.102954$ versus $0.216582$) and 2-Wasserstein distance ($0.103142$ versus $0.216585$) than BE. Furthermore, BGL is closest to the origin in the Wasserstein target diagram, with the smallest 2-Wasserstein distance (0.103142) among the fitted distributions, followed by BE and BW, with BW having a distance of $0.219459$. GEV and LogN have the largest 2-Wasserstein distances ($1.086999$ and $0.756842$). This means that BGL has the best performance in the wind speed distribution fitting among all candidate distributions, which also aligns with the results in \cite{yang2025beta}.

\section{Discussion and Conclusion}\label{sec:discussion}

In this paper, we lift the target diagram to the probability distribution level and propose the Wasserstein target diagram to compare the probability distributions of different models. The original target diagram evaluates errors between paired data of equal sample sizes, and shows the spatial or temporal matching between predicted and observed data. The Wasserstein target diagram instead provides a visual tool to compare probability distributions, and extends the applicability of target diagrams to settings involving unpaired samples, samples of different sizes, and probability distribution fitting to only observed data. Therefore, the choice between the target diagram and Wasserstein target diagram depends on the evaluation objective. Furthermore, Theorem \ref{thm:coincidence} shows that the target diagram and the Wasserstein target diagram coincide if and only if the given data are comonotone, which is often seen for cumulative climate variables indexed by time. 

The two simulation studies in Section \ref{sec:simulation} illustrate the Wasserstein target diagram for samples with equal and unequal sizes. The equal-size simulation shows that the Wasserstein target diagram can capture distributional similarities between independent samples drawn from the same distribution, which have a small Wasserstein distance but a large RMSE. The unequal-size simulation demonstrates the probability comparisons of Wasserstein target diagram without a one-to-one correspondence between data. The real data application on wind speed further illustrates how the diagram distinguishes fitted distributions with small biases.

Note that the Wasserstein target diagram uses the bias, the sign of size (variability), and the centered $2$-Wasserstein distance from the decomposition of 2-Wasserstein distance in Equation ~\ref{eqn:w2_decompostition}. This means that the magnitude of the size and the shape of the distributions are hidden in the Wasserstein target diagram.
A natural modification is to include distribution size or shape information in the Wasserstein target diagram via using different sizes or face colors of the markers.

\section*{Data Availability}
The code and data used in this paper can be found at the following link \url{https://github.com/tiantiy/Wasserstein-Target-Diagram}.

\bibliographystyle{plain} 
\bibliography{refs}

@misc{M2data,
 author = {Jager, David and Andreas, Afshin},
  title = {M2 Meteorology Tower Dataset},
  howpublished = "\url{https://midcdmz.nlr.gov/apps/sitehome.pl?site=NWTC}",
  note = "[Online; accessed 15-November-2026]"
}

@article{chen2026taylor,
  title={Taylor Diagram and {W}asserstein Distance for Model Evaluation},
  author={Chen, Dongwei and King, Emily J and Yu, Hungjui},
  journal={arXiv preprint arXiv:2609.17950},
  year={2026}
}

@article{taylor2001summarizing,
  title={Summarizing multiple aspects of model performance in a single diagram},
  author={Taylor, Karl E},
  journal={Journal of geophysical research: atmospheres},
  volume={106},
  number={D7},
  pages={7183--7192},
  year={2001},
  publisher={Wiley Online Library}
}

@article{jolliff2009summary,
  title={Summary diagrams for coupled hydrodynamic-ecosystem model skill assessment},
  author={Jolliff, Jason K and Kindle, John C and Shulman, Igor and Penta, Bradley and Friedrichs, Marjorie AM and Helber, Robert and Arnone, Robert A},
  journal={Journal of Marine Systems},
  volume={76},
  number={1-2},
  pages={64--82},
  year={2009},
  publisher={Elsevier}
}

@techreport{jager1996nrel,
  title={{NREL} {N}ational {W}ind {T}echnology {C}enter ({NWTC}): {M}2 {T}ower; {B}oulder, {C}olorado ({D}ata)},
  author={Jager, David and Andreas, Afshin},
  year={1996},
  institution={National Renewable Energy Lab (NREL), Golden, CO, United States}
}

@article{irpino2015linear,
  title={Linear regression for numeric symbolic variables: a least squares approach based on {W}asserstein {D}istance},
  author={Irpino, Antonio and Verde, Rosanna},
  journal={Advances in Data Analysis and Classification},
  volume={9},
  number={1},
  pages={81--106},
  year={2015},
  publisher={Springer}
}

@article{irpino2015basic,
  title={Basic statistics for distributional symbolic variables: a new metric-based approach},
  author={Irpino, Antonio and Verde, Rosanna},
  journal={Advances in Data Analysis and Classification},
  volume={9},
  number={2},
  pages={143--175},
  year={2015},
  publisher={Springer}
}

@article{irpino2007optimal,
  title={Optimal histogram representation of large data sets: Fisher vs piecewise linear approximations},
  author={Irpino, Antonio and Romano, Elvira},
  journal={Revue des nouvelles technologies de l'information},
  volume={1},
  pages={99--110},
  year={2007}
}

@article{yang2025beta,
  title={Beta-generalized {L}indley distribution: A novel probability model for wind speed},
  author={Yang, Tiantian and Chen, Dongwei},
  journal={Renewable Energy},
  pages={124216},
  year={2025},
  publisher={Elsevier}
}

@book{figalli2021invitation,
  title={An Invitation to Optimal Transport, {W}asserstein Distances, and Gradient Flows},
  author={Figalli, Alessio and Glaudo, Federico},
  year={2021},
  series={EMS Textbooks in Mathematics},
  number={22},
  address={Berlin, Germany},
  publisher={EMS Press}
}
\end{document}